\newcommand{\ignore}[1]{}
\def\papertitle{On the Impact of Ground Sound}
\def\paperauthorA{Ante Qu}
\def\paperauthorB{Doug L. James}

\documentclass[twoside,a4paper]{article}
\usepackage{dafx_19}
\usepackage{amsmath,amssymb,amsfonts,amsthm}
\usepackage{euscript}
\usepackage[latin1]{inputenc}
\usepackage[T1]{fontenc}
\usepackage{ifpdf}
\usepackage{slashbox}
\usepackage[table]{xcolor}
\usepackage{gensymb}
\usepackage[english]{babel}
\usepackage{caption}
\usepackage{subfig} 
\usepackage{color}
\usepackage{nicefrac}

\ninept

\usepackage{times}
\renewcommand{\vec}[1]{\mathbf{#1}}

\def\orcid#1{\unskip\ignorespaces}

\newif\ifpdf
\ifx\pdfoutput\relax
\else
   \ifcase\pdfoutput
      \pdffalse
   \else
      \pdftrue
\fi

\ifpdf 
  \usepackage[pdftex,
    pdftitle={\papertitle},
    pdfauthor={\paperauthorA, \paperauthorB},
    pdfkeywords={sound synthesis, ground sound, acoustic shader, physical modeling, acoustics},
    colorlinks=false, 
    bookmarksnumbered, 
    pdfstartview=XYZ 
  ]{hyperref}
  \usepackage[pdftex]{graphicx}
\else 
  \usepackage[dvips]{epsfig,graphicx}
  \usepackage[dvips,
    colorlinks=false, 
    bookmarksnumbered, 
    pdfstartview=XYZ 
  ]{hyperref}
  
\fi
  \usepackage[hypcap=true]{caption}
\title{\papertitle}

\affiliation{\paperauthorA \, and \paperauthorB}{Computer Science, \\ Stanford University \\ Stanford, CA 94305 \\ {\tt {[antequ|djames]@cs.stanford.edu}} \orcid{0000-0002-1635-0721}}

\begin{document}
\ifpdf 
  \DeclareGraphicsExtensions{.png,.jpg,.pdf}
\else  
  \DeclareGraphicsExtensions{.eps}
\fi

\maketitle

\begin{abstract}
Rigid-body impact sound synthesis methods often omit the ground sound. In this paper we analyze an idealized ground-sound model based on an elastodynamic halfspace, and use it to identify scenarios wherein ground sound is perceptually relevant versus when it is masked by the impacting object's modal sound or transient acceleration noise. Our analytical model gives a smooth, closed-form expression for ground surface acceleration, which we can then use in the Rayleigh integral or in an ``acoustic shader'' for a finite-difference time-domain wave simulation. We find that when modal sound is inaudible, ground sound is audible in scenarios where a dense object impacts a soft ground and scenarios where the impact point has a low elevation angle to the listening point.
\end{abstract}

\section{Introduction}

Many sound synthesis examples in computer animation and virtual environments contain moving objects that impact the ground or other large flat surfaces. The ground affects the sound in two ways: 1) as a passive scatterer: sound waves in the room are reflected off the ground, and 2) as an emitter: the surface of the ground vibrates due to impact events, and thus emits sound. Typical approaches incorporate the passive scattering and reflection depending on context and methodology; however, very few physics-based approaches consider the acoustic emissions of the ground itself. In this paper we model the ground as an idealized elastodynamic halfspace, and analyze its sound emission during an object-ground impact. Its relative importance is assessed in various object-ground impact scenarios, and is found to vary greatly.

Ground emission and scattering have been explored in many works over the decades. One line of works \cite{cook2002modeling,turchet2010physically} fit data-driven models to synthesize footstep sounds. Works on fracture and micro-collisions \cite{zheng2010rigid,zheng2011toward} treat the ground and table as a large modal vibration source; of these, one paper \cite{zheng2010rigid} models the modes from a 9~m $\times$ 9~m $\times$ 0.9~m concrete slab; these dimensions directly affect modal resonant frequencies. The modal method also requires heavy precomputation resources and storage because large objects have many vibration modes within audible frequencies. Furthermore, the above methods \cite{zheng2010rigid,zheng2011toward} compute propagation with only one object at a time and omit repeated object-ground reflections. 

After an object-ground collision, we may hear three types of sounds: (1) the object emits ringing sound from on its resonant modes, (2) the object emits a transient acceleration noise upon impact, and (3) the ground emits a transient sound upon impact. While many previous papers \cite{zheng2010rigid,zheng2011toward,nishiguchi2018modeling} model the first two in their sound synthesis, they omit the third type of sound. Most recently, \cite{nishiguchi2018modeling} models the collisions of many floor materials; however, it incorporates the floor properties only in the excitation force profile and models sound just from the object's surface. 

One argument for omitting the ground sound is that object sounds are often louder, especially for larger objects, and can mask quieter ground contributions. Nevertheless, depending on the flooring materials, contact parameters, and listening angle, the ground can sometimes be a more efficient sound source than a small object. The interference from the object's reflection can also change its waveform and make it distinct from the ground sound. Our ground vibration model allows us to quantify the intensity of the ground sound, albeit for a simplified elastic halfspace ground model.

Another work \cite{issanchou2017modal} develops a discretized modal model to collide vibrating strings with solid obstacles. We aim not to model ringing sound but rather to introduce a closed-form formula for the transient surface vibrations due to a single impulse. 

Finite-difference, time-domain (FDTD), wave-based sound synthesis methods \cite{raghuvanshi2009efficient,bilbao2015num,wang2018toward} naturally handle scattering with static or moving objects. Recently,  \cite{wang2018toward} enabled scattering with moving objects by rasterizing their boundaries during each timestep. This method abstracts away object sources using an ``acoustic shader'' interface; the simulation queries the object shader for the surface vibrations and uses them to drive sound waves that propagate to the listener. However, no method is proposed to evaluate ground vibrations in an acoustic shader. We implement a ground shader in this work. Our work focuses on the topic of sound emission rather than reflections; it is orthogonal to room acoustics models that simulate room impulse responses and modes.

This surface vibration problem has been studied in seismology literature as Lamb's problem \cite{lamb1904propagation}, and its ideal solution is well-known with a closed form. However, the ideal solution to an instantaneous load contains singularities at wavefronts that are difficult to evaluate numerically. To smooth the singularities, we derive a closed-form temporal regularization of the solution to Lamb's problem that removes the singularities at the three wavefronts, similar to how \cite{de2018dynamic} regularizes the singularity in an infinite elastic medium for animation effects. This closed-form expression makes it easy to model ground sound without simulation. 

{\em We consider the following problem:} Given a simple solid object, such as a ball, colliding with the ground (modeled as an elastic half-space) how do we estimate the sound emitted by the surface vibrations of the ground? Our contributions are 
\begin{enumerate}
    \item an estimate of the material properties and object sizes where the ground sound is not masked by the object sound,
    \item an interactive method to synthesize ground sound (no precomputation is required), and
    \item an ``acoustic shader'' for finite-difference time-domain simulations that directly evaluates the regularized solution.
\end{enumerate}

\section{Ground Sound Model: Background}

We model the transient ground sound by first modeling the ground surface vibration, and then using this motion to drive sound propagation into the air. For the former, we derive a closed-form model of the ground vibration to minimize computation while preserving accuracy. Our propagation model is one-way coupled because air pressure oscillations are not powerful enough to affect the ground.

In particular, we use Lamb's problem \cite{lamb1904propagation} and its solutions to model the floor surface vibrations from an impact, and we describe them in Sections \ref{sec:lambproblem} and \ref{sec:singularities}. We regularize the model in Section~\ref{sec:regularization} to eliminate undesired singularities, and then we model the sound propagation in Section~\ref{sec:applications}.

\subsection{Lamb's problem}
\label{sec:lambproblem}

We present Lamb's problem here, which involves applying an instantaneous normal point load to an elastic halfspace. We present it with a load rather than an impulse in order to simplify the mathematical representation of the solution. In later sections we will derive and use a closed-form representation of the surface acceleration in response to a specific impulse profile.

\begin{figure}[ht]
\centering
\includegraphics[scale=0.75]{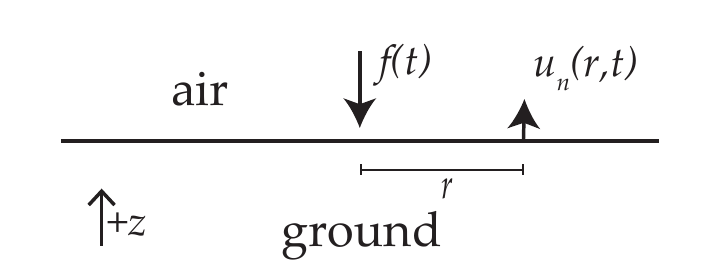}
\caption{{\bf Notation for Lamb's problem:} $f(t)$ is the ground excitation force, and $u_n(r,t)$ is the vertical displacement response. Note that while our diagram shows $f(t)$ in its usual downward direction ($-z$), we define $f(t)$ in \eqref{eq:lambforce} to point in the $+z$ direction.}
\label{fig:lambproblem}
\end{figure}

Consider a linear isotropic elastic half-space with Poisson's ratio $\nu$ and stiffness (shear modulus) $\mu$, as shown in Figure~\ref{fig:lambproblem}. We consider the
elastic half-space to be on the bottom ($z$ negative), and free space to be above it, with the boundary being
the horizontal $z=0$ plane. Starting at time $t=0$, a normal point force of magnitude $1$ is applied and held (``push'') at the origin $(0,0,0)$. The input force profile on the $z=0$ plane is therefore
\begin{equation}
    f(x,y,t) = \; \delta(x,y) \; \theta(t) \; \mathbf{\hat z},
    \label{eq:lambforce}
\end{equation}
where $\delta, \theta$ are the Dirac delta and Heaviside theta functions.

The linear partial differential equations and boundary conditions can be found, for example, in equations 4 and 1 (respectively) of \cite{pekeris1955seismic}; we present their closed-form solution in the next section.

\subsection{Solution to Lamb's problem}
\label{sec:lambproblem}
Pekeris \cite{pekeris1955seismic} first solved Lamb's problem in 1955 for $\nu = \nicefrac{1}{4}$. Others \cite{mooney1974some} later solved it for generic $\nu$. We present the solution for generic $\nu$ from \cite{kausel2013lamb}. Some relevant notation is the following:
\begin{align*}
    c_p &= \text{speed of compression (P)-waves in the medium},\\
    c_s &= \text{speed of shear (S)-waves in the medium},\\
    a &=  \frac{c_s}{c_p} =  \sqrt{\frac{1 - 2\nu}{2 - 2\nu}}, \;\; \;\; r = \sqrt{x^2 + y^2}.
\end{align*}
Define $ \kappa_1^2, \kappa_2^2, \kappa_3^2$ as the complex roots to the Rayleigh equation:
\begin{align}
    16(1-a^2) \kappa^6 - 8 (3-2a^2)\kappa^4 + 8 \kappa^2 - 1 &= 0.
\end{align}
This equation admits three real solutions when $\nu < 0.2631$; otherwise, it has one real root and two complex conjugates. Let $\kappa_1^2$ be the largest real root, and define $\gamma =\kappa_1$. Treat these roots as mathematical tools to help express the result with no direct physical meaning (except that $\gamma$ is the ratio of the S- and R- wave speeds).

\begin{figure}[!htb]
  \includegraphics[width=\linewidth]{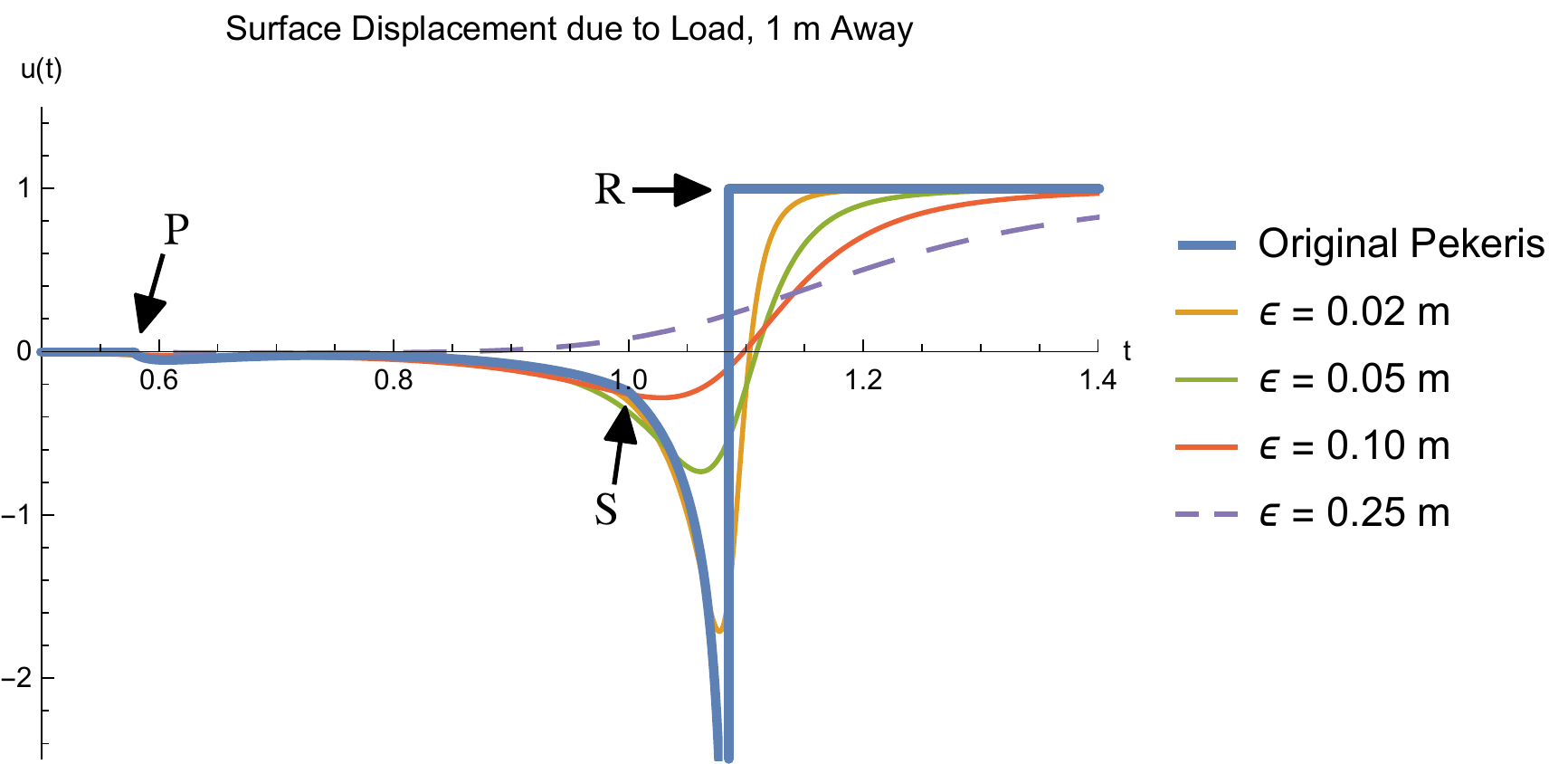}
  \caption{{\bf Elastic wavefronts in time:} (Blue:) Scaled displacement response in the Pekeris solution, at 1~m away. The three wavefronts (P-, S-, R-) are labeled. (Other colors:)  Our temporal regularization, described in Section~\ref{sec:regularization}. The horizontal axis is time in seconds; the vertical axis is scaled normal displacement.}
  \label{fig:fourthloadresponse}
\end{figure}

Define the following set of coefficients:
\begin{align*}
    A_j &= \frac{(\kappa_j^2 - \frac{1}{2})^2 \sqrt{a^2-\kappa_j^2}}{(\kappa_j^2-\kappa_i^2)(\kappa_j^2-\kappa_k^2)}, i \ne j \ne k
\end{align*}
While the response contains both horizontal and vertical displacement, only the vertical motion produces sound. The final vertical displacement response $u_n(r,t)$ is the following:
\begin{align}
    u_n(r,t) &= \frac{1 - \nu}{2 \pi \mu r} \begin{cases} 
      0 & \tau \le a, \\
      \frac{1}{2} \left( 1 - \sum_{j=1}^3 \frac{A_j}{\sqrt{\tau^2 - \kappa_j^2 }} \right), & a< \tau < 1, \\
      1 - \frac{A_1}{\sqrt{\tau^2 - \gamma^2}}, & 1 \le \tau < \gamma , \\
      1 & \tau \ge \gamma ,
   \end{cases}\\
   \tau &= \frac{c_s t } {r}.
\end{align}
This solution applies for all $\nu$, from $0$ to $0.5$ (see \cite{kausel2013lamb}). The piecewise boundaries correspond to the three wavefronts: the pressure P-wave arrives first, when $\tau = a$, travelling at speed $c_p$. The shear S-wave arrives when $\tau = 1$, travelling at speed $c_s$. Finally, the Rayleigh R-wave arrives when $\tau = \gamma$, travelling the slowest at speed $c_r = \nicefrac{c_s}{\gamma}$. See the blue line in Figure~\ref{fig:fourthloadresponse} for an illustration.

Note: It is often convenient to flip the signs of $a^2$, $\gamma^2$, and $\tau^2$ in the square roots of both the numerator and denominator in the terms containing $A_1$, so that the inside of the square root is real. 

\subsection{Singularities}
\label{sec:singularities}
In order to radiate sound waves we need to evaluate the acceleration in the impulse response of Lamb's problem. Unfortunately, the push-like load's displacement response, $u_n(r,t)$, already contains four singularity locations, which means that at each singularity it will be difficult to numerically approximate surface motion.
\begin{itemize}
\item One singularity occurs at all positive $t$ at the origin, where $r = 0$. This singularity occurs due to the spatial $\delta$ load location, and it has asymptotic behavior $\nicefrac{1}{r}$.
\item One singularity occurs at each of the wavefronts---one each at the P- ($\tau = a$), the S- ($\tau = 1$), and the R- ($\tau = \gamma$) wave fronts. The first two wavefronts have continuous but not differentiable singularities. The third wavefront is discontinuous with asymptotic behavior $(\gamma - \tau)^{-\nicefrac{1}{2}}$. 
\end{itemize}

Our goal is to design a regularizing function in time or in space, as a smooth approximation of a delta impulse, to act as the initial force. We then convolve our function with the $u_n$ solution to get a closed-form response that removes the singularities.

We consider the physical parameters of our problem in choosing temporal versus spatial regularization. We would like the regularization parameter to directly match the contact timescale and area. Typical contact radii are much smaller than the contact timescale multiplied by any of the three wave speeds; see Table~\ref{tab:validationparams} for one example. Therefore the spatial contact area smooths the resulting wave by very little compared to the temporal smoothing. Temporal regularization thus gives us a more accurate response than spatial.

\section{Temporal Regularization of the Ground Vibration Model}
\label{sec:regularization}

Consider a function $f_\epsilon(t)$ that approximates $\delta(t)$ on a smoothing timescale $\epsilon$. Since the elastic wave equation is linear and $u_n$ is the response to a Heaviside $\theta$ load, we can get the vertical displacement response to the force, $f_\epsilon * \theta$ (which is an approximate $\theta$), by computing the convolution $u_\epsilon = f_\epsilon * u_n$, or
\begin{align}
    u_\epsilon(r, t) &=  \int_{-\infty}^{\infty} f_\epsilon(t - t' ) u_n(r, t') dt'.
\end{align}
The above gives us the displacement response to a ``push load'' (c.f. \cite{de2018dynamic}). We want an impulse response corresponding to a $f_\epsilon(t)$ force profile. Since $\delta$ is the derivative of $\theta$ and $f_\epsilon$ can be written $f_\epsilon * \delta$, we can subsequently compute the displacement response $w_\epsilon$ to an $f_\epsilon$ impulse force by taking a time derivative of $u_\epsilon$, and likewise the acceleration $a_\epsilon$ by taking more derivatives:
\begin{align}
    w_\epsilon(r, t) &=   \frac{\partial u_\epsilon}{\partial t },\\
    a_\epsilon(r, t) &=  \frac{\partial ^3 u_{\epsilon}}{\partial t ^3}.
\end{align}

We use the regularization function $f_\epsilon$ defined by
\begin{align}
    g_\epsilon(t) &= \frac{ c_s \epsilon}{ \pi( c_s ^2 t^2 +  \epsilon^2)};\\
    f_\epsilon(t) &= 2 g_\epsilon(t) - g_{2 \epsilon} (t).
\end{align}
\begin{figure}[!htb]
  \includegraphics[width=\linewidth]{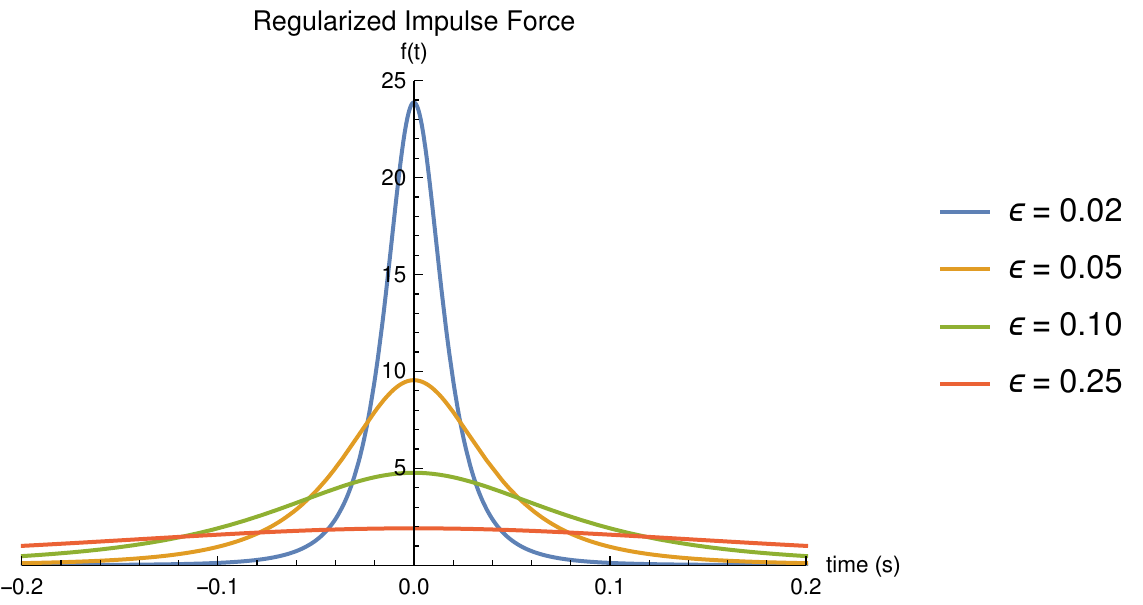}
  \caption{{\bf Smoothed delta function used as the impulse force profile} $f_\epsilon(t)$. Here $\epsilon$ is in meters, the horizontal axis is time in seconds, and the vertical axis is scaled force.}
  \label{fig:fourthforce}
\end{figure}
We chose this function for several reasons. Firstly, it approximates a $\delta(t)$ function as $\epsilon \to 0$: for all $\epsilon$, the total impulse applied is $1$, and as $\epsilon$ gets smaller, a larger proportion of the impulse is applied over a smaller amount of time ($\int_{|t| < \sqrt{\epsilon}} f_\epsilon(t) \to 1$ as $\epsilon \to 0$);
see Figure~\ref{fig:fourthforce} for an illustration. Secondly, it is a smooth approximation of a Hertzian half-sine contact acceleration profile, with timescale $4 \epsilon/c_s$ (see Section~\ref{sec:forceprofile}). Thirdly, while $g_\epsilon$ is only second-order ($g_\epsilon(t) = O(t^{-2})$ as $t \to \infty$), we can form linear combinations of $g_\epsilon$ with varying $\epsilon$ to achieve higher-order falloff, just like the multiscale extrapolation in \cite{de2017regularized}; in this case, our $f_\epsilon$ achieves fourth-order falloff ($O(t^{-4})$).

The final reason is that we can analytically derive the closed-form expression for $u_\epsilon(r,t)$ that is provided in \eqref{eq:regularization} of the appendix. Our regularization eliminates the three singularities at the wavefronts and leaves an integrable, fixed $\nicefrac{1}{r}$ singularity at the origin. In the supplemental material\footnotemark[1] we show that there are no branch cut crossings (a common type of numerical artifact in complex functions) when $\nu \le 0.2631$. We still observe branch cut issues when $\nu > 0.2631$, which is when $\kappa_2^2, \kappa_3^2$ become complex. We recommend using a piecewise polynomial regularization function (see Conclusion Section~\ref{sec:future}) to deal with the branch cuts.

\section{Sound Synthesis}
\label{sec:applications}

\subsection{Impulse profile approximation}
\label{sec:forceprofile}

Similar to \cite{nishiguchi2018modeling,chadwick2012precomputed}, we model the acceleration $a(t)$ using the Hertz contact model. To avoid a discontinuous jerk we approximate the half-sine force with our fourth-order temporal kernel, with $\epsilon/c_s$ set to one-fourth the contact timescale $t_c$:
\begin{align}
    f(t) &\approx J f_\epsilon(t),\\
   4 \epsilon = c_s t_c &= 2.87 c_s \left( \frac{m^2}{a_0 E^{*2}v_n}\right)^{\nicefrac{1}{5}},
\end{align}
where $a_0, m, E^*, J, v_n$ are the object's local radius of curvature, mass, effective stiffness, impulse, and normal impact velocity.

\subsection{Direct sound synthesis via Rayleigh integration}
\label{sec:rayleigh}
Assuming no scattering or absorption from nearby objects, the Rayleigh integral \cite{blackstock2001fundamentals} says the sound pressure at a point $(\vec{r},z)$ due to the plane vibration source is equal to
\begin{align}
    p(\vec{r}, z, t) &= \rho_0 \int_{\mathbb{R}^2} \frac{a_\epsilon(\vec{r}',t-R'/c_0)}{2 \pi R'} d\vec{r}',
\end{align}
where $R' = \sqrt{|\vec{r} - \vec{r'}|^2 + z^2}$, $\rho_0$ is air density, and $c_0$ is the speed of sound in air.

We evaluate this integral numerically in Wolfram Mathematica. We found that the singularity at the origin ($r=0$), mentioned in Section~\ref{sec:singularities}, does not cause issues: to check, we experimented with modified versions of $u_\epsilon$ where in each version we subtract out a ramp $R(r)$ of radius $H$ times the singularity and add back in a ramp $C R(r)$ scaled to have the same average value (from analytically integrating about the origin), and we found that numerically the results were identical to those from the unmodified $u_\epsilon$. We tested radii of $H=0.01$~m, 0.02~m, and 0.10~m. \ignore{We initialize an output buffer for each listening point. We then discretize the acceleration function over a uniform grid and, at each timestep $t$, directly add the contribution of each grid point into each output buffer at the output sample closest to the delayed time $t + R'/c_0$. For a scenario with a grid of size $N_r$ with $N_L$ listening points and impulses lasting a total of $N_S$ samples, the total runtime is $N_S N_r N_L T$ where $T$ is the time it takes to evaluate our acceleration at a point. If the listening points are located directly above the impact point, we can take advantage of cylindrical symmetry and greatly reduce $N_r$. Todo: describe example runtime. $N_r = 6400, N_S < 1000, N_L = 4$. On 36 threads this takes a few seconds. Great speedups are possible with adaptive sampling as future work.}

\subsection{Floor sound shader for FDTD acoustic wavesolvers}

We implemented our floor acceleration model in a general-purpose wavesolver \cite{wang2018toward} that incorporates the scattering of nearby objects. It solves the acoustic wave equation with Neumann boundary conditions
\begin{align}
    \frac{\partial^2 p(\vec x, t)}{c_0^2 \partial t^2} &=  \nabla^2 p(\vec x, t) + \frac{\alpha}{c_0}\nabla^2 \frac{\partial}{\partial t} p(\vec x, t) , & x \in \Omega;\\
    \partial_n p(\vec x , t) &= - \rho_0 a_n (\vec x , t), &x \in \partial \Omega,
\end{align}
by discretizing a region of space onto a rectangular grid and timestepping it with finite differences (see \cite{wang2018toward} for details); here $\Omega$ is the air region, $\partial \Omega$ is the boundary with objects, the subscript $n$ indicates the normal direction, and we set the air viscosity damping coefficient $\alpha=2$E-6~m. The wavesolver samples the boundary normal acceleration $a_n(\vec x , t)$ through acoustic shaders. 

We implemented the floor acceleration model as an ``acoustic shader'' which evaluates the regularized acceleration $a_f(r,t)$ due to each contact impulse, where $r$ is the distance, projected onto the ground plane, between the shader's sample point $\vec{x}$ and the floor impact location. Since there is theoretically an object in contact at the contact point and therefore no adjacent fluid cells, we do not evaluate an acceleration there; therefore the singularity at the contact point ($r=0$) does not cause a problem.

For consistency, we modified the acceleration shader in \cite{wang2018toward} to use the same smooth force profile and impulse evaluation constraints as our ground shader. This also corrects for any amplitude or spectral mismatches between acceleration noise and ground sound. 

\section{Results}
Sound samples for our results are available online.\footnote{\href{http://graphics.stanford.edu/papers/ground/}{http://graphics.stanford.edu/papers/ground/}}

\subsection{Model Validation}

The push-like volume displacement $D$ is given by
\begin{align}
  D(t) &= \int_{\mathbb{R}^2} u_\epsilon(\vec{r},t) \, d\vec{r}.
\end{align}
We evaluate this on a scenario with a small stainless steel ball dropped onto a medium density fiberboard ground and make sure that the volume displacement is consistent with the unregularized Pekeris solution. Relevant parameters are given in Table~\ref{tab:validationparams}. 

\begin{table}[]
    \centering
    \begin{tabular}{|c|c| }
    \hline
         \textbf{Parameter} & \textbf{Value}  \\
         \hline\hline
         Ball Material & Stainless Steel (see Table~\ref{tab:materialparams})  \\
         \hline
         Ground Material & Wood (see Table~\ref{tab:materialparams})  \\
         \hline
         Ball Diameter ($2a_0$) & 2~cm  \\
         \hline
         Drop Distance & 15~cm  \\
         \hline
         Restitution Coefficient ($\kappa$)& 0.5  \\
         \hline
         Impact Location & (0,~0,~0)~m  \\
         \hline
         Listening Location ($\vec{R}$) & (0,~0,~0.2)~m\\
         \hline
         $c_s$ & 2422~m/s \\
         \hline
         Contact Time ($t_c$) & 1.633E-4~s  \\ 
         \hline
         Contact Radius ($r_c$) & 6.316E-4~m \\
         \hline
         $\epsilon =  c_s t_c/4 $ & 9.888E-2~m \\
         \hline
    \end{tabular}
    \caption{{\bf ``Ball Drop'' Simulation Parameters:}
      Scenario information for the validation, the steel ball, wood ground example in Figure~\ref{fig:ballwood}, and the comparisons in Table~\ref{tab:materialcomp}. The lowest frequency nontorsional vibration mode for the steel ball is at 131~kHz, so we omit modal sound. Note that $\epsilon$ is much larger than the contact radius $r_c$, implying that temporal regularization has a much larger smoothing effect than spatial. These parameters are used in the rest of the results unless stated otherwise.}
    \label{tab:validationparams}
\end{table}

\begin{figure}[t]
  \includegraphics[width=\linewidth]{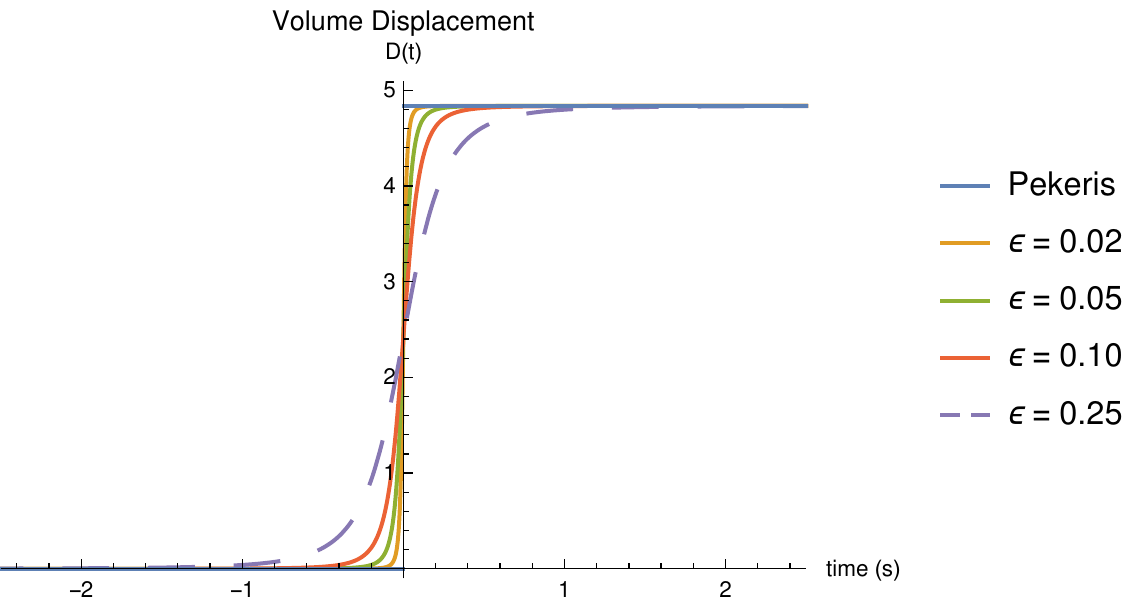}
  \caption{{\bf Volume displacement, $D(t)$:} Here $\epsilon$ is in meters, and the vertical axis is volume displacement scaled by the same factor as in Figure~\ref{fig:fourthloadresponse}. The modified temporal regularization with a smoothed origin proposed in Section~\ref{sec:rayleigh} has a volume displacement plot that looks identical. }
  \label{fig:fourthdisp}
\end{figure}

\label{sec:validatevoldisplacement}
We examined the response to a push load with our temporal regularization. Figure~\ref{fig:fourthloadresponse} plots the vertical displacement at a point 1~m away, and Figure~\ref{fig:fourthdisp} plots the total volume displacement. The curves converge to the Pekeris solution as $\epsilon$ decreases, and asymptotically, each $D(t)$ converges to the correct value as $t \to \infty$.

We also examined the volume displacement, the volume flux, and the momentum flux in response to an impulse. These are each defined as integrating $w_\epsilon$, $ \nicefrac{dw_\epsilon}{dt}$, and $a_\epsilon$ over the $\mathbb{R}^2$ plane. As expected, their curves look like the derivatives of those in Figure~\ref{fig:fourthdisp}.

\subsection{Sound Synthesis Results}
\subsubsection{FDTD Synthesis Examples}
\begin{figure}[h]
\centering
  \includegraphics[width=\linewidth]{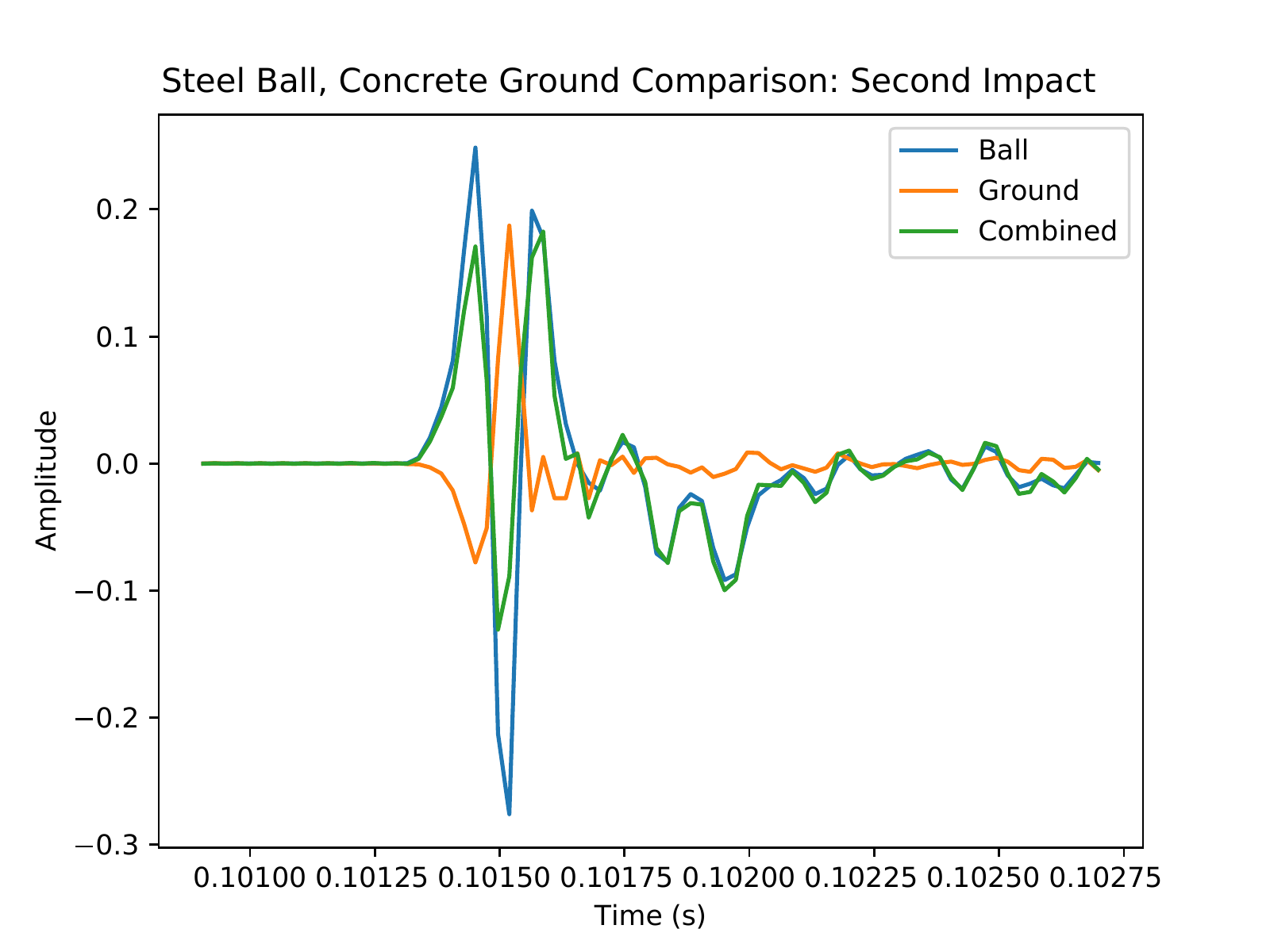}
  \caption[]{An example with 13 balls dropped from various heights onto a concrete ground, simulated with our wavesolver. See the supplemental material for the sound. Each sound (ball, ground, combined) is normalized to 10 Pa. The listening point is at (0.20,~0.12,~0.16)~m, with the $z$ coordinate specifying the height.}
  \label{fig:manyballsconcrete}
\end{figure}

\begin{figure}[h]
\centering
  \includegraphics[width=\linewidth]{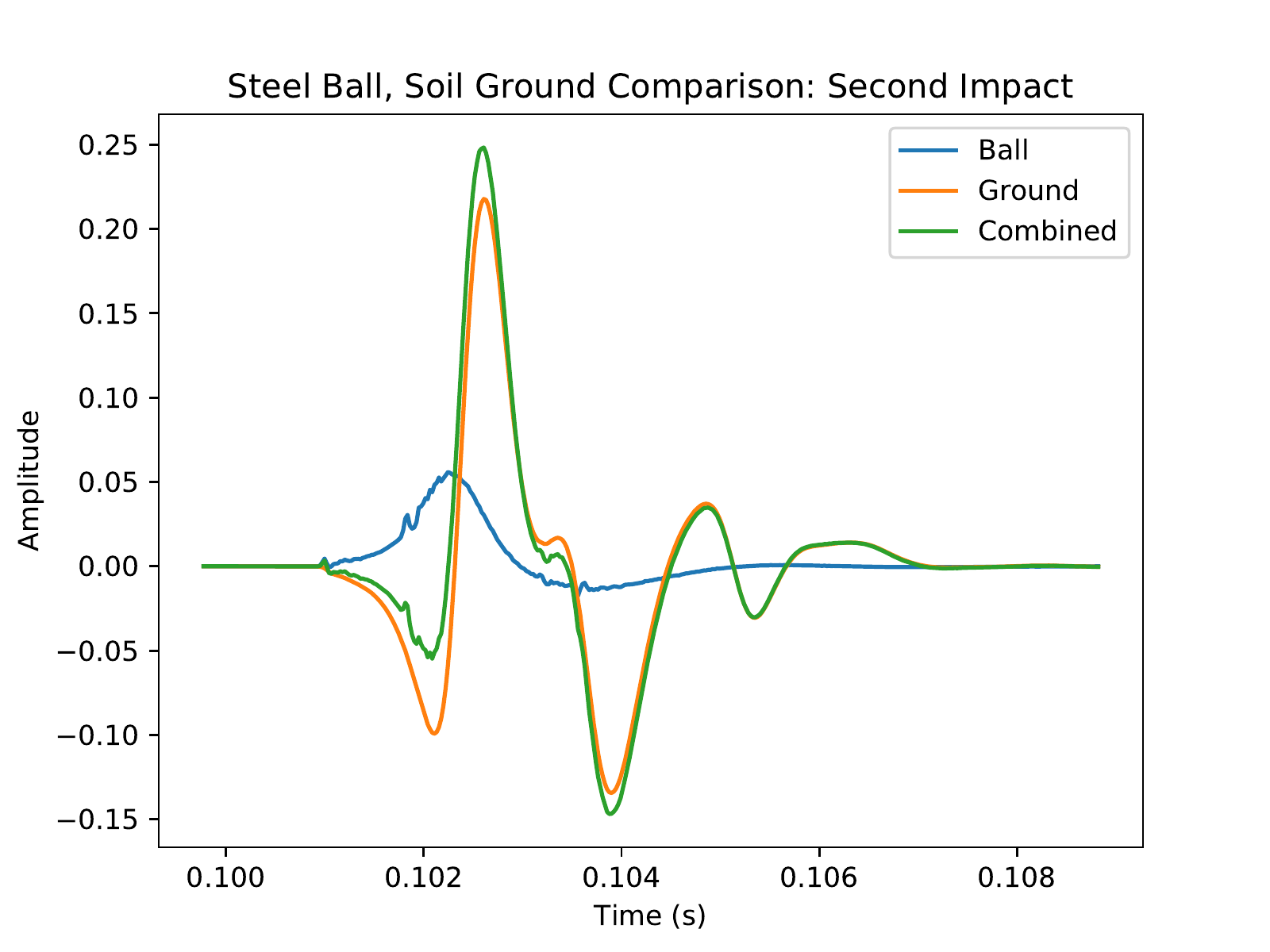}
  \caption[]{An example with 13 balls dropped from various heights onto soil ground, simulated with our wavesolver. See the supplemental material for the sound. Each sound (ball, ground, combined) is normalized to 4.5 Pa. The listening point is at (0.20,~0.12,~0.16)~m.}
  \label{fig:manyballssoil}
\end{figure}

\begin{figure}[h]
\centering
  \includegraphics[width=\linewidth]{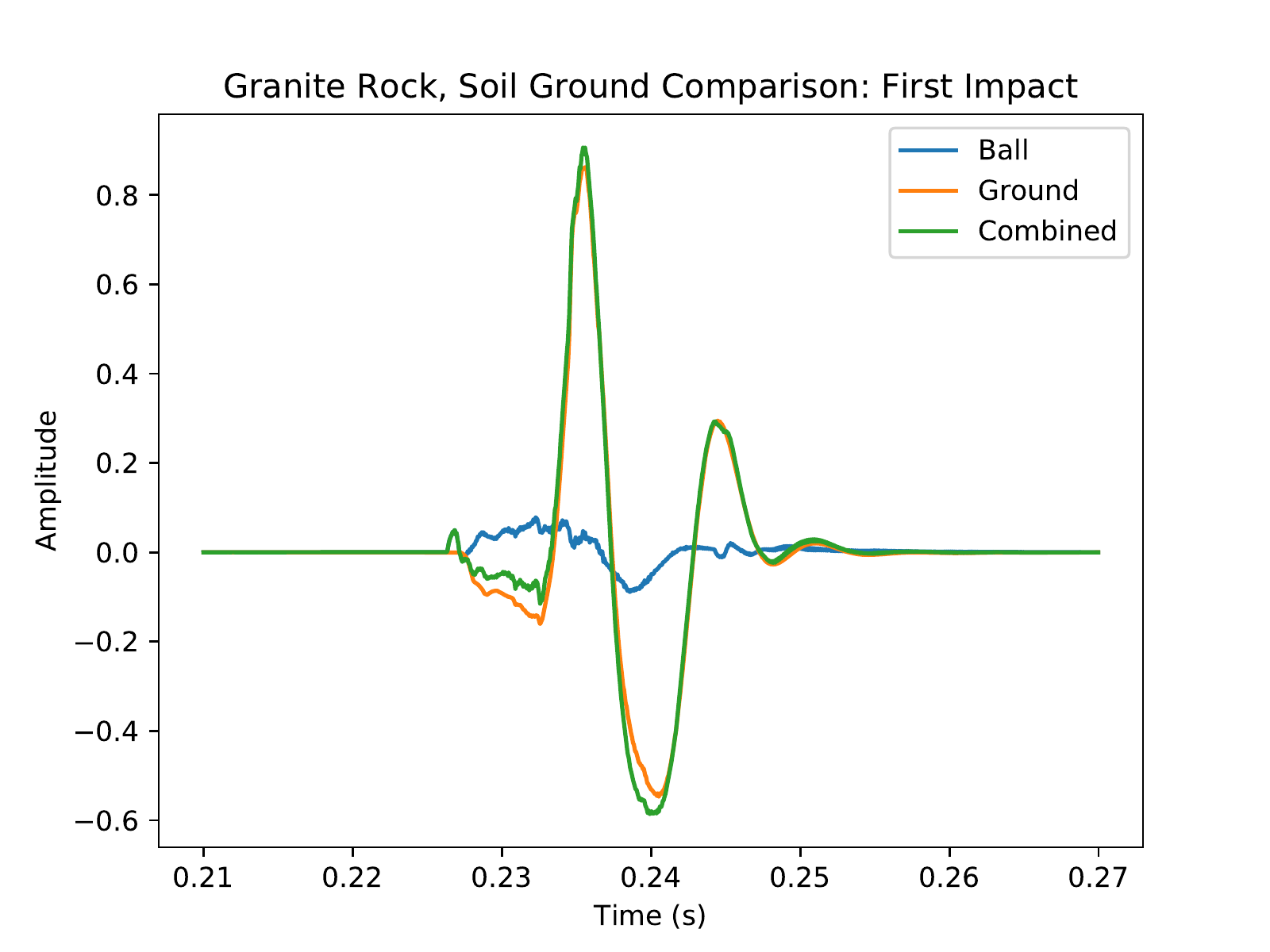}
  \caption[]{An example with a 30~cm spherical granite rock dropped from 25~cm above ground onto soil, simulated with our wavesolver. See the supplemental material for the sound. Each sound (rock, ground, combined) is normalized to 20 Pa. For the almost-silent modal component, we used the modal shader used in \cite{wang2018toward} with Rayleigh damping parameters $\alpha = 6$, $\beta = 1$E-7, in SI units. The listening point is at (0.45,~0.27,~0.48)~m.}
  \label{fig:graniteground}
\end{figure}

We added our ground surface acceleration shader to the time domain simulation system from \cite{wang2018toward}. We also use the modal shader and the acceleration noise shader, which synthesize impact sound for objects. We show a few notable examples in Figures \ref{fig:manyballsconcrete}, \ref{fig:manyballssoil}, and \ref{fig:graniteground}. In each example the modal sound is almost inaudible.

Figure~\ref{fig:manyballsconcrete} shows 13 steel balls with a 2~cm diameter hitting a concrete ground from various heights between 3~cm and 23~cm above ground, and Figure~\ref{fig:manyballssoil} shows these balls hitting a soil ground. Each ball has no audible ringing modes. In both examples the sound from the acceleration noise and the ground have similar frequency spectra. The concrete ground smooths the total sound of the steel ball collision; however, the short duration of the transient sound makes it difficult to discern the sound spectrum. On the other hand, the soil greatly amplifies the total sound from the steel ball collision. Since the ball-soil collision has a longer timescale than the ball-concrete collision, we can hear that the soil sound has a slightly different shape than the ball sound, making the ground relevant.

Figure~\ref{fig:graniteground} shows a spherical granite rock with a 30~cm diameter dropped from a height of 25~cm above ground (centroid at 40~cm). The only audible ringing modes are at much higher frequencies than the contact timescale, hence they were soft, with a peak amplitude of 0.106~Pa. In comparison, the acceleration noise was at a peak amplitude of 1.76~Pa, and the ground contributed a noticeable rumble peaking at 17.2~Pa.

\subsubsection{Ball ground impact comparisons}
Similar to prior work \cite{chadwick2012precomputed}, we can use a closed-form expression to model the sound from a small ball. We treat it as a compact translating sphere, which forms an acoustic dipole source. The far-field acoustic pressure depends on the jerk, with a $\nicefrac{1}{r}$ falloff. The nearfield pressure depends on the acceleration with a $\nicefrac{1}{r^2}$ falloff.
The final expression, according to Eq (6.20) in \cite{rienstra2004introduction}, is
\begin{align}
    p(\vec r, t) &= \frac{\rho_0 a_0^3 \cos(\theta)}{2} \left( -\frac{ a(t - \frac{r-a_0}{c_0} )}{r^2} +  \frac{ \frac{da}{dt}(t - \frac{r-a_0}{c_0} )}{c_0 r}\right)
\end{align}
where $a(t)$ is the acceleration of the ball at time $t$ and $\theta$ is the angle between the acceleration and $\vec r$. We assume perfect reflection and model it by adding the reflection image source of this ball, reflecting the dipole direction and position over the $y$ axis. The total is a longitudinal quadrupole source for hard reflective grounds, and a dipole source for absorptive grounds.

We model the acceleration with the same fourth-order temporal force as that used for the ground in section~\ref{sec:forceprofile}. 
\begin{align}
    a(t) &= -f(t)/ m.
\end{align}
 We simply use $(1+\kappa) m v_n$ as the impulse, where $\kappa $ is the coefficient of restitution of the collision.

\begin{figure}[h]
  \includegraphics[width=\linewidth]{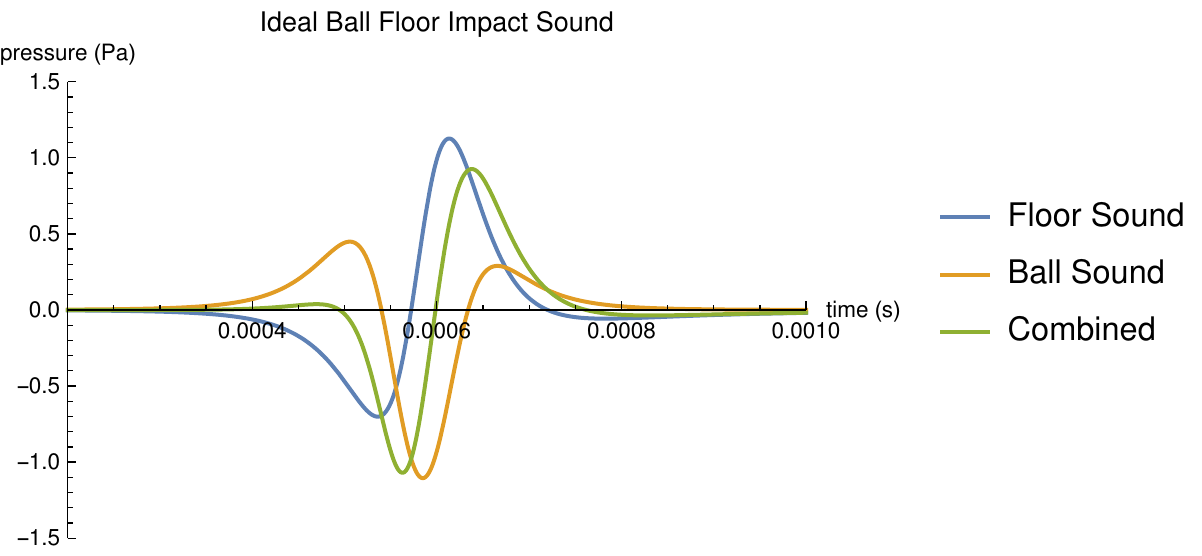}
  \caption{Ideal unobstructed sound for a 2~cm steel ball dropped from 15~cm impacting a wood ground with restitution coefficient 0.5. The listening point is 20~cm directly above the impact point. The quadrupole shape of the ball sound is different from the ground sound, but at high frequencies the frequency content sounds similar and it is hard to tell perceptually. The ground sound adds a significant amount of amplitude to the combined sound, and the combined sound seems to be higher pitched than either sound.}
  \label{fig:ballwood}
\end{figure}

Figure~\ref{fig:ballwood} illustrates an ideal 2~cm steel ball, wood ground impact, with their respective amplitudes. We verified the amplitudes from our wavesolver against these amplitudes. For harder ground materials such as concrete, or lighter object materials such as ceramic, wood, or dice, the ground sound would be much softer compared to the ball sound. The next section generalizes this observation.

\subsection{Impact Sound Parameter Dependence}
\begin{table}[t]
    \centering
    Material Property Reference\\
    \begin{tabular}{|c|c|c|c|}
    \hline
         Material & $E$ (Pa) &  $\nu$ &  $\rho$ (kg~m$^{-3}$) \\
      \hline   
         Stainless Steel & 1.965E+11 & 0.27 & 7955  \\
         \hline    
         Ceramics & 7.2E+10 & 0.19 & 2700  \\
         \hline   
         Granite & 5.07E+10 & 0.28 & 2670  \\
         \hline     
         Concrete & 1.85E+10 & 0.20 & 2250  \\
         \hline    
         Wood & 1.1E+10 & 0.25 & 750  \\
         \hline     
         Plastic (ABS) & 1.4E+9 & 0.35 & 1070  \\
         \hline    
         Soil & 4.0E+7 & 0.25 & 1350  \\
         \hline     
         Paraffin Wax & 5.57E+7 & 0.37 & 786  \\
         \hline
    \end{tabular}
    \caption{{\bf Material parameters used for common materials:} The Young's modulus is $E$, Poisson's ratio is $\nu$, and density is $\rho$. 
    We used medium density fiberboard for wood. }
    \label{tab:materialparams}
\end{table}

\begin{table*}[ht]
    \centering
    Relative Intensities (dB) of Ground Sound Compared to Ball Sound\\
    \begin{tabular}{|c|c|c|c|c|c|c|c|c|}
    \hline
         \backslashbox{ball}{ground} & Steel &  Ceramics & Granite & Concrete &  Wood & Plastic & Soil &  Wax  \\
     \hline         
         Steel & -30.25 & -21.30 & -18.94  & \cellcolor{orange!15} -11.83 & \cellcolor{orange!15}-6.12 & \cellcolor{teal!25} 4.15  & \cellcolor{teal!25} 19.06 &\cellcolor{teal!25} 19.58   \\
         \hline    
         Ceramics &-39.63 & -30.69 & -28.33 & -21.22 & -15.51  &\cellcolor{orange!15} -5.23  &\cellcolor{teal!25}  9.68 &  \cellcolor{teal!25} 10.19\\
         \hline   
         Granite & -39.73 &  -30.78 & -28.43  & -21.32  & -15.60 & \cellcolor{orange!15}-5.33 & \cellcolor{teal!25} 9.58 & \cellcolor{teal!25} 10.10  \\
         \hline     
         Concrete & -41.21 & -32.27 & -29.91  & -22.80 & -17.09 & \cellcolor{orange!15}-6.81 &\cellcolor{teal!25}  8.09 & \cellcolor{teal!25} 8.61  \\
         \hline    
         Wood & -50.76 & -41.81 & -39.46 & -32.34 & -26.63 & -16.36 & \cellcolor{orange!15} -1.45 & \cellcolor{orange!15} -0.93  \\
         \hline     
         Plastic & -47.67 & -38.73 & -36.37 & -29.26 & -23.55 & -13.27 &\cellcolor{teal!25}  1.64 &  \cellcolor{teal!25} 2.15 \\
         \hline    
         Soil & -45.65 & -36.71 & -34.35 & -27.24 & -21.53  & \cellcolor{orange!15}-11.25 & \cellcolor{teal!25} 3.65 &  \cellcolor{teal!25} 4.17\\
         \hline    
          Wax& -50.35 & -41.41 & -39.05 & -31.94 & -26.22  & -15.95 &\cellcolor{orange!15} -1.04 & \cellcolor{orange!15} -0.53 \\
         \hline
    \end{tabular}
    \caption{{\bf Theoretical relative intensity (dB) of ground to ball sound}, for the scenario in Table~\ref{tab:validationparams}. Ball materials are listed on the left, ground on the top. Positive values indicate the ground was louder than the ball. Impact timescale was kept constant at 1.63E-4~s and Poisson's ratio at $0.25$, as neither significantly affect relative amplitude. Scenarios with louder ground sound ($\ge 0$~dB) are highlighted in \colorbox{teal!25}{teal}, and scenarios where the ground sound can be audible (above the most sensitive \textsc{jnd} level of -13~dB \cite{long201481}) are highlighted in \colorbox{orange!15}{light orange}. Note that our overhead listening point is near the maximum relative loudness for the ball, whereas low listening angles tend to receive more ground sound (Figure~\ref{fig:anglecomparison} expands on this relationship).}
    \label{tab:materialcomp}
\end{table*}
Let us describe the impact scenario with the parameters ($t_c$, $a_0$, $v_n$, $\kappa$, $E_f$, $\nu_f$, $c_s$, $\rho_b$,  $R$, $\theta$), where the subscript $f$ indicates ground, $b$ indicates ball, and $(R, \theta)$ indicate the listening point distance and elevation angle. We hereby fix all parameters to their Table~\ref{tab:validationparams} values and vary just one or two of them at a time.

$\rho_b,E_f$: By algebra, the ground sound amplitude is proportional to $\rho_b/E_f$, while the ball sound stays constant. Table~\ref{tab:materialparams} lists these properties for common materials, and Table~\ref{tab:materialcomp} lists the intensity ratio for each material pair. 

$\nu_f$: We found that changing the ground Poisson's ratio does not significantly affect either sound amplitude.

\begin{figure}[h]
  \includegraphics[width=\linewidth]{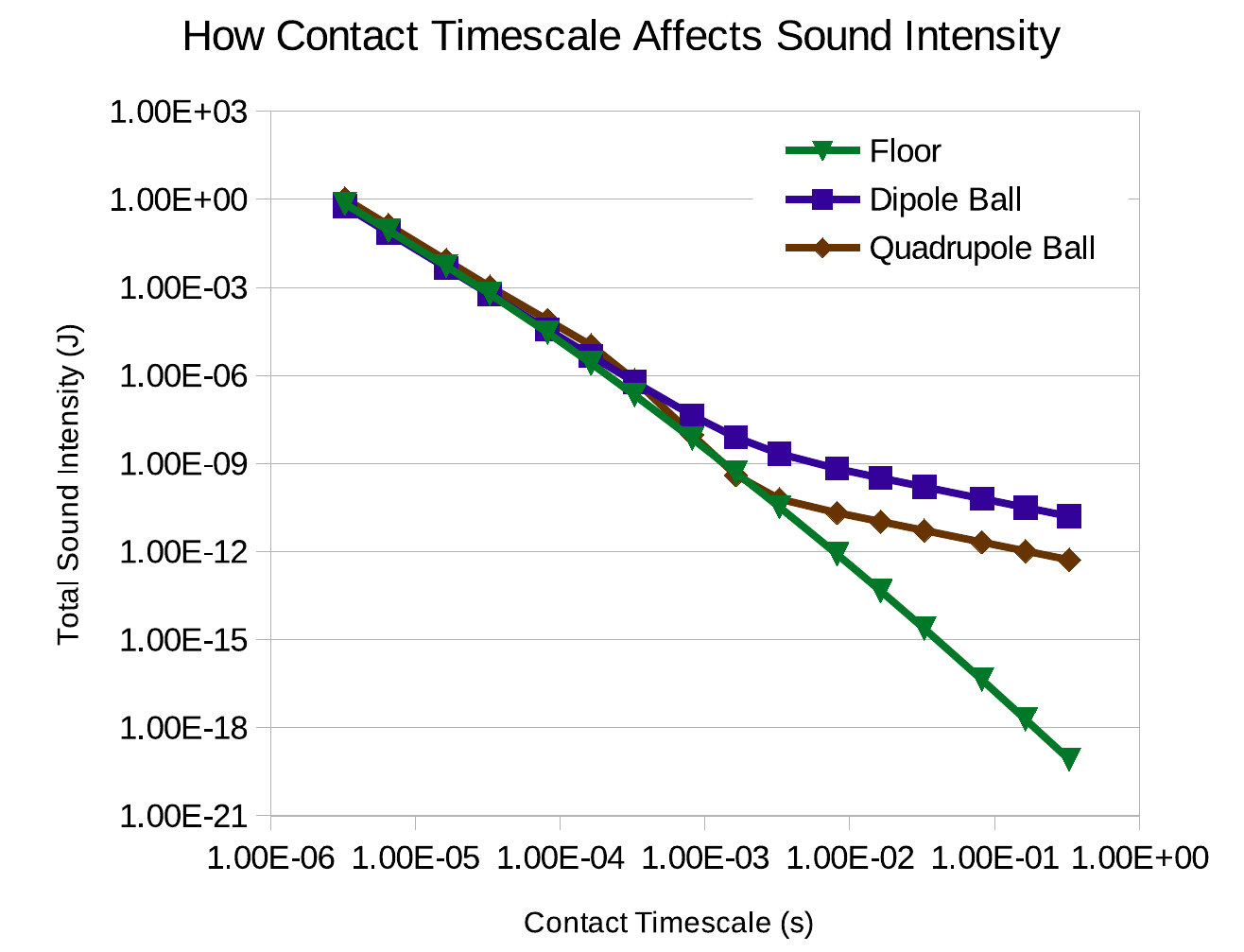}
  \caption{{\bf Sound dependence on contact timescales} measured overhead at $z=20$~cm. For low timescales, the ball and ground both have a $\tau^{-3}$ dependence; however, at high timescales, the near-field term of the ball sound dominates, and its power falls off as $\tau^{-1}$}
  \label{fig:contacttimescale}
\end{figure}

$t_c$: Figure~\ref{fig:contacttimescale} discusses the dependence on contact timescale for one example. In the far field ($R \gg c_0 t_c$) both the ground and the ball sound intensity have similar power law dependence. 

\begin{figure}[h]
  \includegraphics[width=\linewidth]{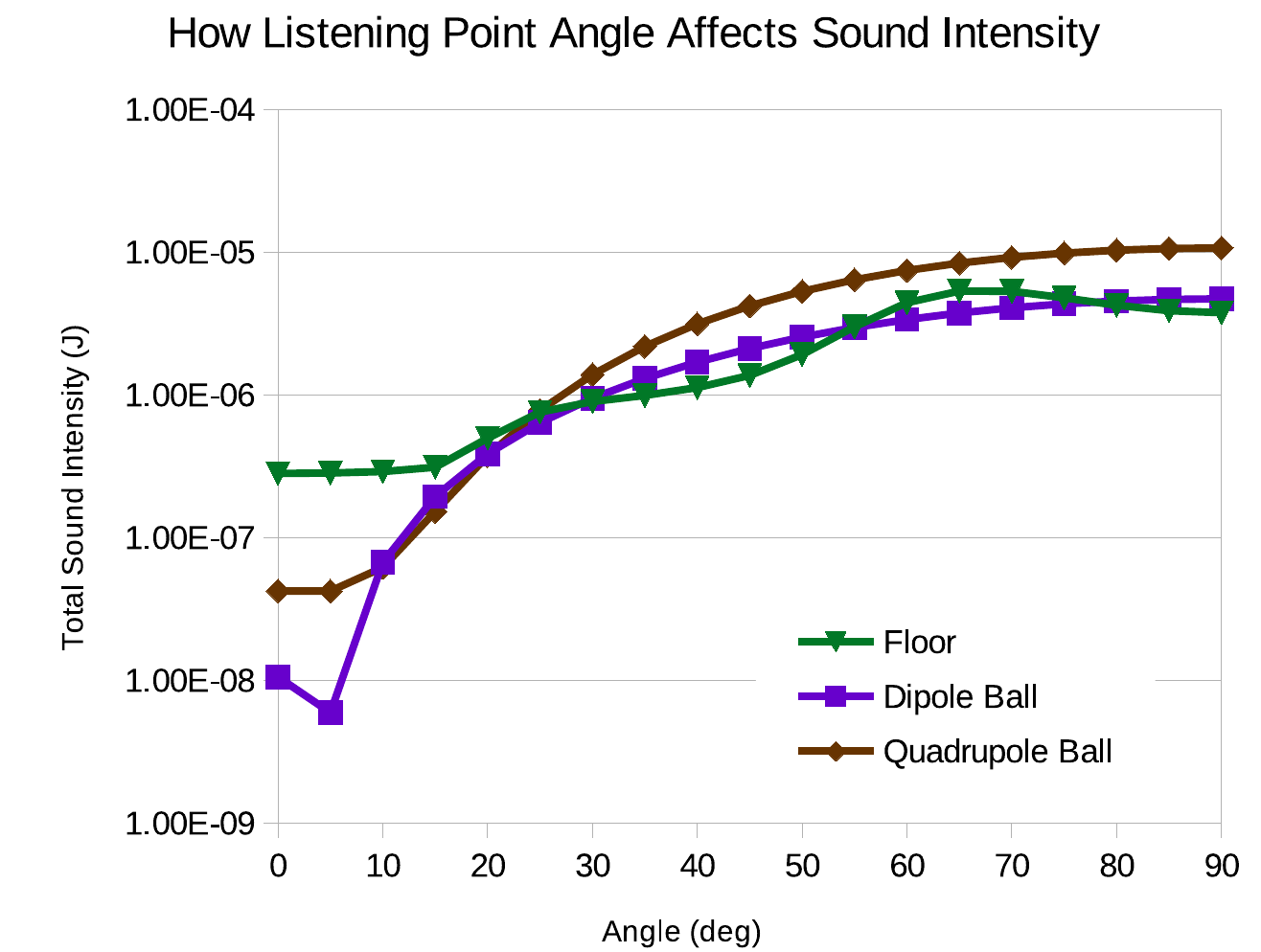}
  \caption{{\bf Angular Dependence of Sound Intensity:} The listening point is 20~cm away, with 90\degree being directly overhead, and 0\degree in the plane. The dipole ball model has a minimum at 5\degree ~because its center is 1~cm above the ground. {\em Observe that the ground sound is significantly louder than the ball sound at low angles.}}
  \label{fig:anglecomparison}
\end{figure}

$\theta$: Figure~\ref{fig:anglecomparison} shows the dependence on listening point angle from the plane. As the listening point gets closer to the plane, the ball sound gets softer at a faster rate than the ground sound. 

\begin{figure}[h]
  \includegraphics[width=\linewidth]{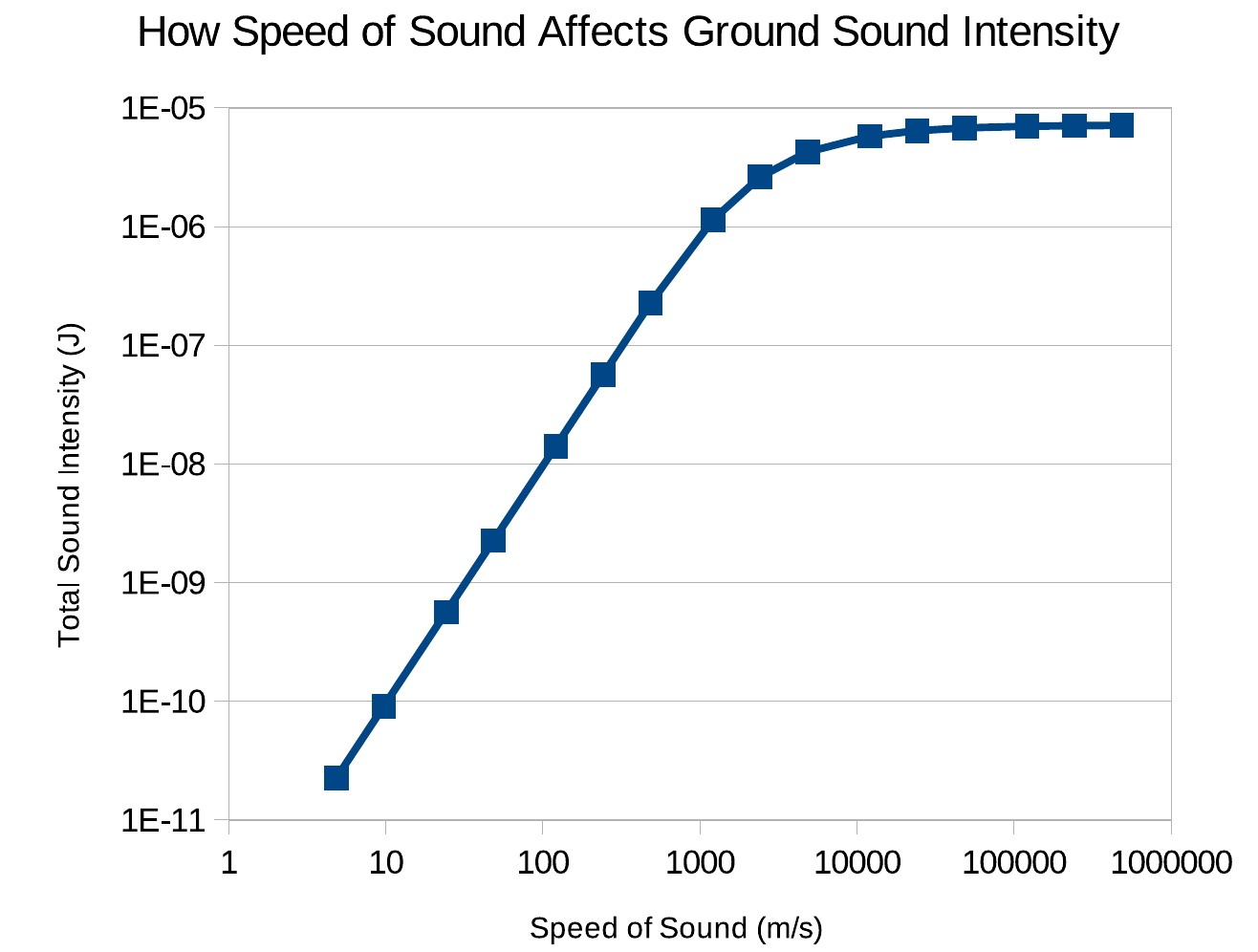}
  \caption{{\bf Ground sound dependence on $c_s$} measured overhead at $z=20$~cm. At low $c_s$, the ground sound intensity is proportional to $c_s^2$, and at high $c_s$, it is constant. The knee cutoff, $c_k$, is about 2576~m/s. By testing a few more parameters, we experimentally determined that $c_k \approx A \sqrt{c_0 R / t_c}$, where $c_0$ is the air speed of sound, $R$ is the listening point distance, $t_c$ is the contact timescale, and $A$ is a dimensionless constant between 3 and 4. }
  \label{fig:speedofsound}
\end{figure}

$c_s$: The ball sound does not depend on $c_s$, the speed of shear waves in the ground, and Figure~\ref{fig:speedofsound} discusses the ground sound dependence on $c_s$. The ground amplitude increases linearly in proportion to $c_s$ until a threshold $c_k \approx A \sqrt{c_0 R / t_c}$ determined by the contact timescale $t_c$ and the listening point distance $R$.

$a_0, v_n, \kappa, R$: In the far field, they affect both sounds equally. 


\subsection{Discussion}
We found that in most everyday scenarios with rigid objects and listening points with high elevation angle, the ground sound would be masked by the object sound: the amplitude of the ball sound is louder, the frequency content is similar, and the contact timescale is often too short to hear the distinct waveforms. In these scenarios, namely the unhighlighted cells in Table~\ref{tab:materialcomp}, we can omit the ground sound.

If the object is dense and the ground has a low shear modulus, then the ground sound can be as loud or louder than the object's acceleration noise. Furthermore, the contact timescale can be slow enough for us to hear the difference between the object and ground sounds. In a few examples we examined, such as steel or granite objects hitting wood, concrete, and soil, the modal ringing sound for the object is too soft, but for larger, less round, and softer objects, the modal ringing sound can dominate the total power output.

\section{Conclusion and Future Work}
We regularized the solution to Lamb's problem to give us a closed-form expression for ground surface acceleration. For impacts from small balls, we used a Rayleigh integral to compute ground sound amplitudes and compared them with object acceleration noise. Furthermore, we implemented an acoustic shader in an FDTD wavesolver to synthesize sound from generic object impacts with the ground, combining modal sound, acceleration noise, and ground sound. We found that the ground sound is more important when the listening point is at a low angle, when the ground has a low shear modulus, or when the object has a high density. Furthermore, ground noise (similar to acceleration noise) is important only for objects where modal ringing noise, which is louder in larger objects, was not audible. In the absence of modal sound, the relative importance of ground sound was not affected by object size in ``ball drop'' tests, notwithstanding changes in contact duration.
\subsection{Limitations and future work}
\label{sec:future}
Our work has several limitations that motivate future work:
\begin{enumerate}
    \item {\em Stable numerical evaluation for general $\nu$:} Our model crosses a discontinuous branch cut when evaluated for high $\nu\ge 0.2631$. We were unable to express the regularized response $u_\epsilon$ in a form that eliminates this branch cut. We explored an alternate regularization using a piecewise polynomial $f_\epsilon = (1 - (t/\epsilon)^2)^n$ for $|t| < \epsilon$, and this gives an expression that does not have the branch jump. However, we used $n=4$ to get a continuous acceleration, and the degree-8 polynomial produces a result that suffers from catastrophic cancellation when $\epsilon$ is small. Future work should ensure stable numerical evaluation for all $\nu$ values.
    \item {\em Finite-depth ground and realistic flooring:} Our ground sound model applies well for ground that is homogeneous for a very deep layer, greater than approximately 50~m deep. For shallower ground layers, the reflections between the layer boundaries form resonance modes that our model does not capture. Furthermore, when an object is dropped onto a hard floor in a building, we hear the vibrational response of the building. Future work could model the responses of more realistic building and flooring structures.
    \item {\em Tangential frictional loads:} We only modeled the vertical response to a vertical load. Future work can regularize the closed-form solutions for a vertical response to a tangential load, such as incurred by contact friction.
    \item {\em No closed-form sound:} We provided an expression for surface acceleration but not the sound. Future work could derive a model for the final sound based on listening position.
\end{enumerate}
\section{Acknowledgements}

   We thank the anonymous reviewers for their feedback. We acknowledge support from the National Science Foundation (NSF) under grant DGE-1656518, the Toyota Research Institute (TRI), and Google Cloud Platform compute resources. Any opinions, findings, conclusions, or recommendations expressed in this material are those of the authors and do not necessarily reflect the views of the NSF, the TRI, any other Toyota entity, or others.

%

\nocite{*}
\bibliographystyle{IEEEbib}
\bibliography{groundsound_DAFx19} 
\appendix
\section{Derivation of Regularized Response}
In this derivation, we let $t' = c_s t$, and we note that at the end, we need to scale by the right power of $c_s$.
\begin{align}
    g_\epsilon(t') &= \frac{\epsilon / \pi } {t'^2  + \epsilon^2 }\ignore{\\
    u_n (r, t') &= \frac{1 - \nu}{2 \pi \mu }\begin{cases} 
      0 & t' \le a r, \\
      \frac{1}{2} \left( \frac{1}{r} - \sum_{j=1}^3 \frac{A_j}{\sqrt{ t'^2 - \kappa_j^2 r^2 }} \right), & a r <  t' < r, \\
      \frac{1}{r} - \frac{A'_1}{\sqrt{  \gamma^2 r^2 -  t'^2}}, & r \le t' < \gamma r , \\
      \frac{1}{r}, & t' \ge \gamma r .
   \end{cases}}
\end{align}
We want to find the convolution $k'_\epsilon=g_\epsilon(t') * u_n(r,t')$. This represents the displacement response to an arctan load, which approximates the Heaviside theta load. Define  $ \mathcal{U}, \mathcal{W}, \mathcal{V},$ as the following
\begin{align}
    \mathcal{U}_\epsilon(t', \sigma) &= \frac{1}{r} \int_\sigma^\infty g_\epsilon (t' - s) ds;\\
    \mathcal{V}_\epsilon(t', s, \alpha) &=  \int \frac{ g_\epsilon (t' - s)}{ \sqrt{s^2 - \alpha^2}} ds ;\\
    \mathcal{W}_\epsilon(t', s, \alpha) &=  \int \frac{g_\epsilon (t' - s)}{ \sqrt{\alpha^2 - s^2} } ds.
\end{align}
Integrating,
\begin{align}
    \mathcal{U}_\epsilon(t', \sigma) &= \frac{1}{2 r} + \frac{1}{\pi r} \arctan\left(\frac{t' - \sigma}{\epsilon}\right);\\
    Z_\epsilon(t', \alpha) &= \sqrt{\alpha^2 + (\epsilon - i t')^2};\\
    \mathcal{V}_\epsilon(t', s, \alpha) &= \text{Re}\left( \frac{1 }{  \pi Z_\epsilon(t', \alpha)} \left( - \log( \epsilon -i ( t' - s) )\vphantom{\sqrt{2}}\right.\right.\nonumber \\
    &\;\;\; \left. \left.+\log( \alpha^2 - (t' + i \epsilon) s - i Z_\epsilon(t', \alpha) \sqrt{s^2 -\alpha^2}) \right) \right);\\
    \mathcal{W}_\epsilon(t', s, \alpha) &= \text{Im}\left( \frac{-1 }{  \pi Z_\epsilon(t', \alpha)} \left( - \log( \epsilon -i ( t' - s) )\vphantom{\sqrt{2}}\right.\right.\nonumber \\
    &\;\;\; \left. \left.+\log( \alpha^2 - (t' + i \epsilon) s + Z_\epsilon(t', \alpha) \sqrt{\alpha^2 -s^2}) \right) \right).
\end{align}
Check the Mathematica notebook on the website\footnotemark[\value{footnote}] for verification.

Plugging in the integration limits, the convolution $k'$ is
\begin{align}
    k'_\epsilon(r, t') &= \frac{1 - \nu}{4 \pi \mu} \left( \mathcal{U}_\epsilon(t',a r) + \mathcal{U}_\epsilon(t', r) \vphantom{\sum_{j=2}^3}\right.\nonumber  \\
    &+ 2\mathcal{W}_\epsilon(t', \gamma r,\gamma r)- \mathcal{W}_\epsilon(t',  r,\gamma r) - \mathcal{W}_\epsilon(t', a r,\gamma r)\nonumber \\
    &+\left.\sum_{j = 2}^3( \mathcal{V}_\epsilon(t',r,\kappa_j r) -  \mathcal{V}_\epsilon(t',a r,\kappa_j r) ) \right).
\end{align}
Our final expression, in terms of the original $t$, is
\begin{align}
    k_\epsilon(r,t) &= k'_\epsilon(r,c_s t),
\end{align}
that is, there is no missing $c_s$ scale factor because the extra $c_s$ from the convolution is cancelled by the missing $c_s$ from normalizing $g_\epsilon$.
For fourth-order, we simply take
\begin{align}
\label{eq:regularization}
    u_\epsilon(r,t) &= 2 k_\epsilon(r,t) -  k_{2\epsilon}(r,t).
\end{align}
In the supplemental material\footnotemark[1] we show that when $\nu \in [0, 0.2631)$, this solution does not cross any branch cuts as we vary $(r,t)$.
\ignore{
\subsection{No branch cut crossings when $0 \le \nu < 0.2631$}
We show here that our solution does not cross a principal branch cut when $0 \le \nu < 0.2631$.

For these $\nu$, $\kappa_j$ are real, and $\kappa_2, \kappa_3 <a$. \cite{kausel2006fundamental} This means $s > \alpha$ in $\mathcal{V}$, $\alpha > s$ in $\mathcal{W}$, and $\epsilon,s,\alpha > 0$. The principal branch cut for both $\sqrt{z}$ and $\log(z)$ are at the negative real line.

$Z_\epsilon$: The radicand of $Z_\epsilon$ never approaches the negative real line: the only way to achieve zero imaginary part is for $t'=0$, when the radicand is a positive real number.

The first log in $\mathcal{V}$ and $\mathcal{W}$: $\epsilon + s - it'$ has a positive real part.

Second log in $\mathcal{V}$: because $Z_\epsilon$ is a square root, it has positive real part. Therefore the entire expression inside the log has negative imaginary part, never crossing the negative real line.

Second log in $\mathcal{W}$: When $t' \ge \alpha^2/s$, the imaginary part of $Z_\epsilon$ is negative, meaning the entire expression inside the log is in the third and fourth quadrants. When $t' < \alpha^2/s$, the real part of the expression inside the log is positive, meaning it is in the first and fourth quadrants. Overall, the expression only lives in the first, third, and fourth quadrants, implying that it cannot cross the negative real line.

We have shown that no branch crossings occur for any pair of $(r,t)$ in our solution.}
\end{document}


\title{On the Impact of Ground Sound: Supplemental}
\author{Ante Qu and Doug L. James}
\date{June 19, 2019}
\maketitle

\section{No branch cut crossings when $0 \le \nu < 0.2631$}
We show here that our regularized solution, detailed in Appendix A, does not cross a principal branch cut when $0 \le \nu < 0.2631$.

For these $\nu$, $\kappa_j$ are real, and $\kappa_2, \kappa_3 <a$ \cite{kausel2006fundamental}. By inspection of (26), this means $s > \alpha$ in $\mathcal{V}_\epsilon$, $\alpha > s$ in $\mathcal{W}_\epsilon$, and $\epsilon,s,\alpha > 0$. The principal branch cut for both $\sqrt{z}$ and $\log(z)$ are at the negative real line.

$Z_\epsilon(t',\alpha)$: The radicand of $Z_\epsilon$ never approaches the negative real line: the only way to achieve zero imaginary part is for $t'=0$, when the radicand is a positive real number.

The first log in $\mathcal{V}_\epsilon(t',s,\alpha)$ and $\mathcal{W}_\epsilon(t',s,\alpha)$: $\epsilon + s - it'$ has a positive real part, so it never crosses the negative real line.

Second log in $\mathcal{V}_\epsilon(t',s,\alpha)$: because $Z_\epsilon$ is a square root, it has positive real part. Therefore the entire expression inside the log has negative imaginary part, never crossing the negative real line.

Second log in $\mathcal{W}_\epsilon(t',s,\alpha)$: When $t' \ge \alpha^2/s$, the imaginary part of $Z_\epsilon$ is negative, meaning the entire expression inside the log is in the third and fourth quadrants. When $t' < \alpha^2/s$, the real part of the expression inside the log is positive, meaning it is in the first and fourth quadrants. Overall, the expression only lives in the first, third, and fourth quadrants, implying that it cannot cross the negative real line.

We have shown that no branch crossings occur for any pair of $(t',s)$ in our solution.
\bibliographystyle{IEEEbib}
\bibliography{supplement} 

\begin{figure}[b]
  \vspace{-0.6cm}
{\scriptsize{\it Copyright:~\copyright \hspace*{1 pt} 2019 Ante Qu and Doug L.~James. This is an open-access article distributed under the terms of the \href{http://creativecommons.org/licenses/by/3.0/}{Creative Commons Attribution 3.0 Unported License}, which permits unrestricted use, distribution, and reproduction in any medium, provided the original author and source are credited.}}
\end{figure}